\documentclass[12pt,preprint]{aastex}

\def\lessim{\mathrel{\hbox{\rlap{\hbox{\lower4pt\hbox{$\sim$}}}\hbox{$<$}}}}
\def\grtsim{\mathrel{\hbox{\rlap{\hbox{\lower4pt\hbox{$\sim$}}}\hbox{$>$}}}}

\shorttitle{Eclipses of TT~Tri}
\shortauthors{Warren, Shafter \& Reed}

\begin{document}


\title{Modeling Eclipses of the Novalike Variable TT Triangulum}


\author{S. R. Warren\footnote{Current Address: Department of Astronomy, University of Minnesota, 116 Church Street S. E., Minneapolis, MN 55455}, A. W. Shafter \& J. K. Reed}
\affil{Department of Astronomy and Mount Laguna Observatory\\
     San Diego State University\\
    San Diego, CA 92182}
\email{warren@astro.umn.edu,aws@nova.sdsu.edu,jreed@sciences.sdsu.edu}




\begin{abstract}
Multicolor ($BVRI$) light curves of the eclipsing novalike variable
TT~Tri are presented.
The eclipse profiles are analyzed with a parameter-fitting model
that assumes four sources
of luminosity: a white dwarf primary star, a main-sequence
secondary star, a flared accretion disk with a rim,
and a bright spot at the intersection
of the mass-transfer stream and the disk periphery. Model parameters include
the temperatures of the white dwarf ($T_1$) and the secondary star ($T_2$),
the radius ($R_\mathrm{d}$) and temperature ($T_\mathrm{d}$)
of the disk periphery, the inner disk radius ($R_\mathrm{in}$),
the disk power-law radial brightness
temperature exponent ($\alpha$) and
thickness parameter ($h_\mathrm{r}$),
and a bright spot temperature enhancement factor ($\chi_\mathrm{s}$).

A grid of model model light curves was computed, covering an extensive range of
plausible parameter values. The models were then compared with the mean
$BVRI$ light curves to determine
the optimum values for the fitting parameters and their associated errors.
The mass ratio of TT~Tri is poorly constrained in our models,
but must lie roughly in the range $0.3\lessim~q(=M_2/M_1)\lessim0.9$.
Models characterized by mass ratios of $q=0.3$ ($i=76.1^{\circ}$),
$q=0.6$ ($i=72.6^{\circ}$), and $q=0.9$ ($i=70.4^{\circ}$) 
were all capable of providing acceptable fits to the data, although the
best fits were achieved for mass ratios near the upper end of the
permitted range ($q=0.6$ and $q=0.9$). 
The values of the remaining fitting parameters were found to be
insensitive to the adopted mass ratio and rim thickness.
The accretion disk was found to extend to $\sim50-60$\% of the distance to
the inner Lagrangian point in all models, but came closer to
reaching the tidal limit (as expected for steady-state accretion)
in the higher mass ratio models. The same behavior was found for the
radial temperature profile of the disk, which increased with mass ratio,
becoming more consistent with that expected for steady state accretion
in the $q=0.6$ and $q=0.9$ models.
Models with a truncated inner disk ($R_{\mathrm{in}}>>R_1$) generally
resulted in a higher white dwarf temperature, and a steepening of the disk
temperature profile, but were not
required to achieve a viable steady-state disk solution.
No evidence was found for a luminous bright spot in the system, which
is not surprising given the lack of a pre-eclipse ``hump" in the
light curve. 

A total of 22 eclipse timings were measured for the system,
which yielded an ephemeris for the times of mid-eclipse
given by ${\rm JD}_{\odot} = 2,453,618.953(3) + 0.1396369(4)~E$.
A comparison of the observed brightness and color at mid-eclipse
with the photometric properties of the best-fitting model
suggests that TT~Tri lies at a distance of $\sim400-500$~pc.

\end{abstract}





\section{Introduction}

TT~Tri (GR 286, HS0129+2933) is a little-studied cataclysmic variable star,
identified originally as an eclipsing binary
by Romano (1978a,b) in a sample of eleven variable stars discovered on
Schmidt plates of a field near M33. Except for the refined astrometric
measurements of Sharov (1992),
the star languished in obscurity until
it surfaced in the Hamburg Quasar Survey (Hagen et al. 1995)
as HS0129+2933. It has subsequently
been suggested to be a member of the SW~Sex subclass of novalike variables
by Rodr\'igues-Gil (2005), who measured an orbital period of 3.35~hr for the
system.

The study of eclipsing cataclysmic variables is valuable
because it provides the opportunity to study the
radiative properties of accretion disks through analysis of the
eclipse profiles.
In an attempt to further elucidate the properties of this long-neglected
system, we have obtained extensive, time-resolved, multi-color CCD photometry
of the eclipses of TT~Tri, and subsequently analyzed the
resulting eclipse profiles with a parameter-fitting eclipse model.
In this paper we present the results of that study.

\section{Observations}

Observations of TT~Tri\footnote{$\alpha=01:31:59.86,
\delta=+29:49:22.1$ (epoch 2000.0);
$\ell=133^{\circ}$, $b=-32^{\circ}$ (Downes et al. 2001).}
were carried out during 14 nights in 2005 September -- December using
the Mount Laguna Observatory 1~m reflector.
On each night a series of
exposures (typically 30~s) were taken through either a Johnson-Cousins
$B,V,R,$ or $I$ filter (see Bessel 1990), and imaged on a Loral $2048^2$ CCD.
To increase the time-sampling efficiency, only a $400\times400$
subsection of the
full array was read out. The subsection was centered on TT~Tri and
included several nearby stars that were used as comparison objects for
differential photometry.
A summary of observations
is presented in Table~1.

The data were processed in a standard fashion (bias subtraction and
flat-fielding) using IRAF.\footnote{
IRAF (Image Reduction and Analysis Facility) is distributed by the
National Optical Astronomy Observatories, which are operated by AURA, Inc.,
under cooperative agreement with the National Science Foundation.}
The individual images were subsequently
aligned to a common coordinate system and
magnitudes for TT~Tri
and the two comparison stars were then determined using the
{\it IRAF\/} APPHOT package.
Atmospheric extinction variations were removed to first order by dividing
the flux of TT~Tri by a nearby comparison star.
The $B$ and $V$ differential light curves were then placed on an
absolute scale by calibration of the comparison star against the
standard stars in Landolt (1992).
During the 14 nights of observation, we observed a total of 22 eclipses
-- six in $B$ and $R$, seven in $V$, and three in $I$.
The resulting eclipse light curves are
displayed in Figures~1 and~2,
with the mean photometric parameters given in Table~2.
The uncertainties in the mean magnitudes for the $B$ and $V$ light curves
reflect systematic variations in the disk luminosity (including short-term
``flickering") that presumably results
from fluctuations in the mass transfer rate.
The individual photometric measurements have uncertainties of $\sim0.03$~mag.

\section{TT~Tri Ephemeris}

Despite its identification as an eclipsing binary nearly 30 years ago,
there are no published eclipse timings available for TT~Tri,
so we have used our 22 eclipse timings of TT~Tri to establish one.
Following earlier studies, the
times of mid-eclipse have been determined
by fitting a parabola (a second order polynomial with no cross term)
to the lower half of the eclipse profile and
taking the minimum to be the time of mid-eclipse.
A linear least-squares fit of the resulting mid-eclipse times (see Table~3)
yields the following ephemeris for TT~Tri:

\begin{equation}
T_{\mathrm{mid-eclipse}} = {\rm JD}_{\odot}~2,453,618.953(3)+0.1396369(4)~E.
\end{equation}

Residuals of the individual eclipse timings with respect to this ephemeris
are also given in Table~3, and are plotted as a function of cycle number
in Figure~3. The relatively short ($\sim3$ month) interval spanned
by our observations
does not allow us to assess the long-term stability of the orbital period
of TT~Tri.

\section{The Eclipse Model}

The TT~Tri data have been analyzed using the
latest version of parameter-fitting eclipse modeling program
described in Shafter \& Misselt (2006, hereafter SM06).
The model considers four principal
sources of light from the system: the white dwarf,
the secondary star, the accretion disk (including a rim),
and the bright spot where the
inter-star mass transfer stream impacts the periphery of the disk.
The model light curve flux at a given orbital
phase is computed by summing the contribution from the secondary star
and the unocculted regions of the remaining three sources of radiation.
The fluxes from all four sources of radiation have been corrected
via the use of a linear limb-darkening law with passband-specific
coefficients given in van Hamme (1993).
Although the model fits only the shapes and depths of the eclipse profiles,
we compute a model $B-V$ color from the assumed
blackbody fluxes using the transformations given in Matthews \& Sandage (1963)
that can be compared with the observed $B-V$ color.

As in previous studies, the disk in TT~Tri is assumed to
radiate like an optically-thick blackbody, characterized by
a radial, power-law temperature profile.
A comprehensive mathematical description of the model computations can be found
in SM06.
As a point of clarification, we note that the expressions for
the disk (surface and rim) and the primary and secondary star fluxes
(eqns. 3--8) of SM06 imply that the
specific intensity of the radiating surfaces can be approximated
by the Planck function multiplied by the linear limb darkening law.
However, the Planck function does not represent the
specific intensity normal to the radiating surface, as assumed, but rather
the specific intensity integrated over all emergent angles.
The use of model atmospheres in lieu
of blackbody radiation in future versions of the code
would solve this problem, but in the meantime
the fluxes given in SM06 can be approximately corrected by
dividing the relevant expressions by the factor $1-\mu_{\lambda}/3$ that
results when the linear limb darkening expression is integrated over
$2\pi$ steradians. Fortunately, the omission of this factor did not
affect our earlier results significantly since the eclipse model
uses relative fluxes to compute normalized light curve intensities;
thus, the factors of $1-\mu_{\lambda}/3$ (which are similar for all
sources of light in the system) effectively cancel in the computation
of the model light curves. 

In the case of a steady-state accretion disk,
the relation between the effective temperature and radius is given by
(e.g. Pringle 1981):
\begin{equation}
\sigma T_\mathrm{eff}^4(r) = {3 G M_1\dot M \over 8\pi r^3}~\left[1 - \sqrt{R_\mathrm{in} \over r}\right],
\end{equation}
where $R_\mathrm{in}$ is the inner radius of the disk.
For $r>>R_\mathrm{in}$, eqn. (2) reduces to
the well-known relation $T_\mathrm{eff}(r)\propto~r^{-3/4}$. However,
the temperature we consider in our model is actually a brightness temperature,
not an effective temperature. Thus,
to generalize the temperature dependence with radius, we allow for temperature
gradients that deviate from the steady state value. In previous papers,
we have assumed $T_\mathrm{br}(r)\propto~r^{-\alpha}$, however, this simple parameterization
will overestimate the disk temperature at small radii.
Thus, in our present
study, we employ a modified power law
given by:
\begin{equation}
T_\mathrm{br}(r) = T_{\mathrm{d}} \Bigg({R_{\mathrm{d}} \over r}\Bigg)^{\alpha}~\left[1 - \sqrt{R_\mathrm{in} \over r}\right]^{(\alpha/3)},
\end{equation}
where $T_\mathrm{d}$ is the brightness temperature of the rim at the
outer edge of the disk. Since the disk's
brightness temperature profile (as fit by our model) is not precisely the same
as its effective temperature profile, a steady state disk will not necessarily
be characterized by $\alpha=0.75$.

One of the significant advantages of the inclusion of a disk rim
is that the bright spot can be more easily modeled.
The presence of the disk rim provides both an isotropic and
an anisotropic component to the bright spot radiation.
The isotropic component of the bright spot is produced as in
Shafter et al. (2000).
by radiation from
a region of the disk's surface defined by
the intersection of a circular area of radius $0.2~R_{\mathrm{d}}$
(centered on the point of intersection of the mass-transfer stream
and the disk perimeter) and the accretion disk. The anisotropic
component is produced by radiation from the corresponding
azimuthal region of the disk rim.
As in earlier studies, the bright spot is assumed to radiate like a blackbody,
with the bright spot temperature parameterized
by a multiplicative factor, $\chi_\mathrm{s}$, applied to the
local disk temperature in the bright spot region.

The white dwarf primary star and the secondary star are assumed to radiate as
blackbodies of temperature $T_1$ and $T_2$, respectively, with
the tidally-distorted shape of the Roche-lobe-filling secondary star
fully taken into account. The effects of gravity darkening and
irradiation of the secondary star by the primary component are neglected
in our model since their effects are expected to be minimal in our
analysis. Not only does the secondary star contribute a small fraction
of the system's total light, we only model a limited range of orbital
phase near primary eclipse.
The radii of the white dwarf and
secondary stars are computed from the orbital period and mass ratio
as described in Shafter et al. (2000).

\subsection{Input Parameters}

The input parameters required by the model are summarized in Table~4.
They can be divided into two general categories:
fixed parameters and fitting parameters. The fixed parameters
are those parameters that are either known, a priori, such as the
orbital period, or those whose
values can be estimated using assumptions implicit in the model.
The fitting parameters, on the other hand, cannot be
specified initially, and are varied during the fitting procedure.

Since the orbital period (P=3.35~hr) and eclipse width ($\Delta\phi=0.035$)
are known,
the masses and dimensions of the binary system can be computed
once the mass ratio, $q(=M_2/M_1)$ is specified (e.g. see Shafter et al. 2000).
Unfortunately, the mass ratio of TT~Tri is unknown.
For a given secondary star mass, possible values of the mass ratio
can be constrained both by the requirement that the white dwarf star
not exceed the Chandrashekahar limit, and by the requirement for
stable mass transfer.
The dependence of $M_2$ on $q$
is weak, and to first order the mass of the secondary star can
be expressed as a function of the orbital period alone. From
Warner (1995, eqn. 2.100) we have $M_2/M_{\odot}\simeq0.065P^{5/4}(\mathrm{hr})$;
for an orbital period of 3.35~hr, we therefore estimate $M_2\simeq0.3M_{\odot}$.
Thus, the requirement that $M_1 < M_\mathrm{ch}$ leads to a lower limit of
$q\grtsim0.25$. A firm upper limit on the mass ratio is more difficult to
establish. Models that treat the secondary star as a fully convective
polytrope require
$q\lessim2/3$ for stable mass transfer (e.g. Politano 1988). However,
more recent studies suggest that stable accretion may be possible
at mass ratios approaching unity
(e.g. see Han et al. 2002, and references therein).
Thus, the mass ratio of TT~Tri must lie
in the range $0.25\lessim~q\lessim1.0$.
Given that $M_2\sim0.3$M$_{\odot}$ and the average white dwarf masses
in novalike systems is believed to be of order $0.8~M_{\odot}$
(Smith \& Dhillon 1998), it would appear likely that the actual mass ratio
of TT~Tri lies closer to the lower end of this range.
Nevertheless,
in order to explore the effect of the mass ratio
on our analysis, we have explored solutions for three representative
mass ratios: $q=0.3$, $q=0.6$, and $q=0.9$.
Following the analysis described in Shafter et al. (2000),
the possible masses, dimensions, and orbital inclinations of
the TT~Tri system have been computed for our representative
mass ratios, and are summarized in Table~5.

In addition to the mass ratio, the spectral type and
temperature of the secondary star
can be estimated prior to beginning the fitting procedure.
In their statistical study of the properties of cataclysmic variable
stars, Smith \& Dhillon (1998) found an empirical relation between
the orbital period and spectral type of the secondary star. For an
orbital period of 3.35~hr, the most likely spectral type is
$\sim$M4V. According to Popper (1980) a star with
this spectral type is expected to have a temperature
of $\sim3400$~K. Thus, we have chosen to fix
$T_2=3400$~K, and not vary this parameter during the fitting procedure.
If the secondary star in TT~Tri deviates significantly from the
mean properties of CVs with similar orbital periods and
measured spectral types (e.g. because it deviates from the
main-sequence mass-radius relation), our estimate of $T_2$,
and thus the light contributed by the secondary star, would clearly become
less reliable.

Another input parameter that can be constrained to a limited extent
is the disk rim thickness, $h_{\mathrm{r}} = h/R_{\mathrm{d}}$, where $h$ is the
height of the rim perpendicular to the plane of the disk and
$R_{\mathrm{d}}$ is the disk radius. As discussed in SM06,
studies by Meyer \& Meyer-Hofmeister (1982) and Smak (1992) have shown that the
rim height can be approximated by $h_{\mathrm{r}}\simeq0.038\dot M_{16}^{3/20}$
(Warner 1995; eqn. 2.52).
For plausible mass accretion rates between $\sim10^{16}$~g~s$^{-1}$ and
$\sim10^{18}$~g~s$^{-1}$,
we estimate rim heights in the range of $h_{\mathrm{r}}\sim0.04$ to
$h_{\mathrm{r}}\sim0.08$.
To bracket the range, we consider specific rim heights of both
$h_{\mathrm{r}}=0.04$ and $h_{\mathrm{r}}=0.08$ in our models of TT~Tri.
We chose not to consider a finer grid of possible rim heights given that
the relatively low inclination of the TT~Tri system makes it unlikely
that the precise choice of rim thickness will have a significant impact on the
model results.

The values of the remaining parameters, the
``fitting parameters", which include the radial
disk temperature parameter $\alpha$,
and the temperatures of the disk rim,
bright spot, and component stars, are constrained through the model fit.

\section{Light-Curve Fitting}

For ease of computation, the individual light curves for each color were
averaged together before beginning the fitting procedure. All but one
of the 22 eclipse light curves
were considered for the modeling routine. The initial $V$ light curve from
2006 September 05
had a different time resolution from the rest of the data,
and was used soley in the determination of the
ephemeris.  The remaining light curves were converted to orbital phase, and
placed on a relative intensity scale. The
light curves for each color were then combined by phase-binning
the data ($\Delta\phi = 0.005$) in the phase range $-0.2<\phi<0.2$,
and averaging.  The four resulting
$BVRI$ light curves were then normalized by their out-of-eclipse light levels.
Since the post-eclipse light level is systematically brighter and less stable
than the pre-eclipse level in all colors,
we arbitrarily assigned the pre-eclipse data a relative weight twice
that of the post-eclipse data when computing the normalization.

In previous studies of eclipsing novalike variables,
the brightness temperature profiles of the accretion disks
are often found to be significantly less steep than one would expect
for steady-state accretion (e.g. see
B\'{\i}r\'o 2000, Robinson et al. 1999, Rutten et al. 1992).
Smak (1994) and Knigge et al. (2000) have shown that the addition of
a rim can steepen the derived temperature profile of the disk.
More recently, in their study
of V~Per, SM06 found that the inclusion of a disk
rim, or a truncation of the inner accretion disk (as might arise from
the partial disruption of the disk by the white dwarf's magnetic field),
or both, resulted in best-fitting models with steeper values of the
radial disk temperature exponent, $\alpha$.
In the case of TT~Tri, the shallower eclipses and lower inclination angle
reduces the importance of the disk rim (e.g. Smak 1994). However,
a truncated inner disk might still have a significant effect on the
value of $\alpha$.  To explore this possibility further, we have
computed eclipse models with truncated inner disks.
A value of $R_{\mathrm{in}}=0.2R_{\mathrm{L1}}$, which was adopted by
SM06 in their study of V~Per, proved to be too
large to produce plausible fits to the TT~Tri data. Instead, we considered
two sets of models with less extensive inner disk
disruptions: one set with
$R_{\mathrm{in}}=0.05R_{\mathrm{L1}}$, and the other with
$R_{\mathrm{in}}=0.10R_{\mathrm{L1}}$.

To begin the fitting procedure, we select plausible ranges of values for the
five fitting parameters, indicated in Table~4 as ``variable".
Each parameter is then varied one at a time
through a suitable range while the remaining parameters
are held constant. In this way a 5-dimensional parameter space of
models is explored for plausible solutions.
The goodness of fit at each stage is determined using a standard
$\chi^2$ test.
The deviations between the model and the data are determined for
each of the four colors, and the final $\chi^2$ statistic is
determined by weighting each of the colors equally. Since we model the
light curve near eclipse only, the calculation of $\chi^2$ has been
restricted to orbital phases between $\phi=-$0.08 and $\phi=0.08$,
which corresponds roughly
to the time just prior to the
onset of disk ingress to the completion of bright-spot egress.
The range over which each parameter is varied has been chosen to be
sufficiently large so that all plausible model solutions were explored.
A thorough sampling required eight values to be considered
for each fitting parameter. Thus, a total of 32,768~($=8^5$) models
were computed for each mass ratio and rim thickness considered.
The best-fitting solutions for our grid of models standard disk
(SD) and truncated disk (TD) models are given Tables~6 and~7.

\subsection{The Model Solution Distributions}

As found previously in our studies of other eclipsing CVs,
exploring the matrix of possible solutions reveals that many
combinations of parameters produce good fits to the data.
The goal is to determine which particular combination of parameters
best represents the true model for TT~Tri. The best model fits are
characterized by a reduced chi-square, $\chi_\nu^2\simeq1.1$;
however,
a wide range of parameters give acceptable fits to the
light curves.

To explore the
range of plausible solutions,
frequency distributions for the 100 best-fitting models were
constructed for each of the five fitting parameters,
and for the the model $B-V$ color. As examples, the distributions
for the standard disk models with $h_\mathrm{r}=0.04$ for our three
representative mass ratios are shown in Figures~4--6.
The mean and standard deviation of each parameter distribution, along with
the mean $\chi_\nu^2$ for
all models are summarized in Tables~8 and~9.
As long as the number of model solutions included in the distributions
is sufficiently small so that $<$$\chi_\nu^2$$>$~$\lessim2.0$
(which represents a plausible fit to the data),
the distribution means do not depend strongly
on the precise number of model solutions included in the distributions.
The $<$$\chi_\nu^2$$>$ values suggest that the higher mass ratio ($q=0.6$ and
$q=0.9$) SD and TD models provide better
fits to the data than do the $q=0.3$ models,
although the latter models cannot be unequivocally
ruled out based on their fits alone.

The frequency distributions reveal how well a given parameter
is constrained by the model fit, with
the mean and standard deviation providing
an estimate of the parameter's optimum value, and
a quantitative assessment of its uncertainty.
One of the more tightly constrained parameters is the
disk radius, $R_{\mathrm{d}}$,
where a radius of $\sim40-60$\% of the distance to the inner Lagrangian
point is required by virtually all the models.
The value of this parameter is primarily
determined by the phase width where the disk eclipse begins and ends,
and is relatively insensitive to other system parameters
(e.g. Sulkanen et~al. 1981).
The maximum size of the accretion disk is limited to the tidal
radius, which for TT~Tri is $\sim0.74R_\mathrm{L1}$ for $q=0.3$,
$\sim0.68R_\mathrm{L1}$ for $q=0.6$,
$\sim0.62R_\mathrm{L1}$ for $q=0.9$ (Warner 1995, eqn. 2.61).
Given that the re-distribution of
angular momentum in a steady-state disk is expected to result in
an increase in the disk radius out to near the tidal radius,
it appears once again that the higher mass ratio models, where the disk fills
a greater fraction of the tidal radius, are perhaps more realistic.

Despite the fact that the best-fitting SD models are characterized by
values of $\alpha\simeq0.6$,
the distributions reveal
that viable solutions are possible with the nominal
steady-state value of $\alpha\simeq0.75$.
If we force the
SD models to have $\alpha=0.75$, the $q=0.6$ and $q=0.9$
models provide significantly
better fits to the data than do the $q=0.3$ models, with little change
in the best-fitting values of the other parameters.
For this reason, and because the higher
mass ratio models produce a better fit
to the data and yield a more plausible disk radius, it appears likely
once again that the mass ratio of TT~Tri lies in the upper end of the allowed
range.

Despite the indirect evidence for a mass ratio near the upper
end of the allowed range, our eclipse modeling does not enable a
precise value of the mass ratio in TT~Tri to be determined,
as is evidenced by the fact that both the $q=0.6$ and $q=0.9$ models,
for example, produce nearly identical fits to the data.
Fortunately, as can be seen from
Tables 6 and~7, and from Figures 4--6, the
properties of the accretion disk determined by our model fit
are quite insensitive
to the adopted mass ratio. Given that a mass ratio of $q=0.9$ is near
the limit of mass transfer stability, and would be unusual for a CV
with such a short orbital period,
we consider the $q=0.6$ models to best represent the TT~Tri system.
Figure~7 shows the best-fitting $q=0.6$ SD model
plotted together with the phased eclipse data.

As is typical for a novalike variable,
the light curves do not exhibit any significant
asymmetry of the eclipse profile, and there is no prominent ``hump" prior to
eclipse, so it is not surprising that
none of the models considered required a significant temperature enhancement
from the bright spot in the system. Furthermore, the eclipse profile,
which lacks the sharp transitions on ingress and egress
characteristic of the white dwarf eclipse
suggest that the white dwarf star does not contribute significantly
to the system light (except in the case of the TD models where the
white dwarf luminosity partially compensates for the inner disk radiation).
Thus, as expected, none of the best-fitting SD models required a particularly
luminous (hot) white dwarf.

\section{Discussion}

\subsection{Disk Structure}

The geometry of the TT~Tri system can be visualized by referring
to Figure~8, which shows a pictorial representation of
the system geometry at mideclipse for our $q=0.6$ model.
As suggested by the relatively shallow eclipse depths ($\sim 50$\% of the
out-of-eclipse level), the white dwarf is just barely eclipsed at
$\phi=0$.
Not only does the relatively low orbital inclination for TT~Tri
result in only partial eclipse of the accretion disk, it also
minimizes the contribution of the disk rim to the overall disk light.
Thus, unlike our models for the high-inclination system, V~Per
(SM06),
the thickness of the disk rim did not have a significant
effect on our model solutions.

Although the TD
models are generally characterized by
higher values of the disk temperature parameter, $\alpha$,
viable SD models can be found with steep temperature profiles
($\alpha\simeq0.75$) typical of what is expected for a
steady-state disk. Thus, there is no compelling evidence
for a truncated inner disk in TT~Tri based on our eclipse studies.
Nevertheless,
the system has not yet been extensively studied, and
perhaps future observations
(e.g. polarimetry) may reveal signatures of a magnetic white dwarf in
this system.

\subsection{Correlations Between Model Parameters}

Insights into the effect that the various input parameters
have on acceptable model solutions can be found by exploring correlations
between the parameters. We begin by considering our expanded set of model
solutions defined by $\chi_\nu^2<2.0$, which allows potential correlations
to be studied over an expanded range in parameter space. Correlations between
pairs of parameters are explored by allowing them to vary while
holding the remaining parameters fixed at their optimum values. For a total
of five fitting parameters, there are a total of 10 such pairings. As
a representative example, Figure~9 shows
the correlations based on our best-fitting $q=0.6$ SD model.

For the standard disk models, the most highly correlated parameter
pairs involve $R_{\mathrm{d}}$, $T_{\mathrm{d}}$, and $\alpha$,
which together determine the disk luminosity.
In cases where the accretion disk is completely occulted at mid-eclipse
(e.g. high inclination systems and/or
longer orbital period systems with higher mass ratios),
the relative eclipse depth is determined uniquely by the integrated
disk luminosity.
For a constant disk luminosity, the radial
temperature parameter $\alpha$ should be negatively correlated with the
temperature at the disk periphery,
$T_{\mathrm{d}}$, and the disk radius, $R_{\mathrm{d}}$,
as was indeed observed 
in our earlier studies of GY~Cnc (Shafter et al. 2000)
and EX~Dra (Shafter \& Holland 2003).
However, as shown in our recent study of V~Per,
another system where the disk is never completely obscured,
the eclipse depth no longer uniquely defines the disk luminosity.
The normalized eclipse depth can be achieved with a relatively
low-luminosity
disk with a shallow radial temperature gradient (small $\alpha$),
and a
cooler outer disk temperature, or a smaller outer disk radius, or both.
Alternatively, the required eclipse depth can be achieved with a
more luminous disk characterized by a steeper temperature profile
coupled with a higher outer disk temperature, a larger outer disk
radius, a bright disk rim, or some combination of the three.
As the radiation from the outer disk is increased, the inner disk
luminosity must increase to compensate, requiring a steeper temperature
gradient and a higher value of $\alpha$.
Thus, for a given disk radius and rim thickness,
$\alpha$ should be {\it positively} correlated with the outer
disk (rim) temperature, $T_{\mathrm{d}}$, as was observed in the case of V~Per,
and now for TT~Tri as shown in Figure~9.


\subsection{Potential Model Constraints from the Disk Luminosity}

Although a range of models can produce
acceptable fits to the data, as just discussed
they need not represent the same disk luminosities.
Thus, it is worth exploring whether
the disk luminosity expected for the TT~Tri system
can be used to constrain the set of viable models.
The disk luminosities can be explored through two independent approaches:
directly from our model fluxes, and from an estimate of the mass
accretion rate inferred from our models.

In the first approach, we can
compare the ratio of our model $V$ fluxes for the disk and secondary star
as viewed from an inclination angle of $i=0$ (i.e. ``face-on"). This
quantity, $\eta_\mathrm{d,s}=F_\mathrm{V}^\mathrm{d}(i=0)/F_\mathrm{V}^\mathrm{s}(i=0)$,
is given in Tables~10 and~11 for our best-fitting SD and TD models,
respectively. If we ignore limb-darkening and assume a
spherical secondary star and a flat circular disk, then the luminosity
ratio is approximately equal to
half the model flux ratio (a flat disk has half the surface area
of a sphere). Thus,
$L_\mathrm{V}^\mathrm{d}/L_\mathrm{V}^\mathrm{s}\simeq\eta_\mathrm{d,s}/2$, and
an estimate of the secondary
star's luminosity will yield a crude estimate for the disk luminosity
(neglecting the small contribution of the rim to the overall light.).
As an example, the flux ratio of our best-fitting $q=0.6$ SD model
is $\eta_\mathrm{d,s}=126$,
which represents a luminosity ratio of $\sim$63. Thus,
we estimate the
disk in this model to be $\sim$4.5 mag brighter than the secondary star.
An estimate of $M_\mathrm{V,2}$
can be found given the secondary star's spectral type,
which we estimated earlier to be $\sim$M4V.
For a main-sequence secondary star of radius
$R_2=0.34$~R$_{\odot}$ (Table~5) and spectral type M4V,
we find $M_\mathrm{V,2}\simeq11.5$ (Popper 1980). Thus, for this model,
we estimate an absolute visual magnitude for the accretion disk, and in
effect for the TT~Tri system (the white dwarf and secondary star
contribute little $V$ light compared to the disk),
to be $M_\mathrm{V}\sim7.0$.
This value is roughly $1-1.5$ magnitudes fainter that expected
for a $P=3.35$~hr novalike system (e.g. Warner 1987).
If we consider the truncated inner disk model
($R_\mathrm{in}=0.10R_\mathrm{L1}$) with $\eta_\mathrm{d,s}=131$,
we find a similar value with $M_\mathrm{V}\sim6.9$.

As a check on these results, following SM06
we can also make a rough estimate of the disk luminosity
from an estimate of the mass transfer rate coupled with the $\dot M$ vs.
$M_\mathrm{V}$ relation from Tylenda (1981).
By assuming the disk is in a steady-state, and
adopting the model radius and temperature of the outer disk (rim),
we can estimate the mass accretion rate for a given model from eqn (2).
Specifically,
\begin{equation}
\dot M = {8 \pi R_\mathrm{d}^3 \sigma T_\mathrm{d}^4 \over 3 G M_1}\left[1 - \sqrt{R_\mathrm{in} \over r}\right]^{-1} \mathrm{g~s}^{-1}.
\end{equation}
Mass transfer rates ($\dot M_{17} = \dot M/10^{17}~\mathrm{g~s}^{-1}$)
for our best-fitting models are given in Tables~10 and~11.
These values should be considered rough estimates given that
we are implicitly taking $T_\mathrm{eff}=T_\mathrm{br}$, and neglect
any heating of the outer disk that may be caused by irradiation from
the white dwarf and inner disk.
For the specific case of the $q=0.6$ SD model
we have converted the mass-transfer rate to an equivalent $M_\mathrm{V}$ through
the relation given in Table~1 of Tylenda (1981).\footnote{
A small correction ($\sim0.8$~mag) has been applied to the resulting
absolute magnitude to account for the difference in $M_1$ and $R_\mathrm{d}$
between Tylenda's model ($M_1=1M_{\odot},
R_\mathrm{d}=5\times10^{10}~\mathrm{cm}$) and our model.
Following Warner (1995),
the correction for $R_\mathrm{d}$ has been estimated from the models of
Wade (1984).} We find a value of $M_\mathrm{V}\sim5.8$, which is
somewhat brighter, and more typical of that expected for a novalike variable,
than the value of $M_\mathrm{V}\sim7.0$
estimated earlier from the disk-to-secondary star luminosity ratio.

Since TT~Tri does not exhibit dwarf nova eruptions,
a further constraint on possible models for this system is provided
by the requirement that the mass transfer rate exceed the critical
value for stable accretion, $\dot M_\mathrm{crit}$.
A convenient expression for the critical mass transfer rate
is given by
$\dot M_\mathrm{crit}\simeq10^{16} R_\mathrm{10}^{21/8} M_1^{-7/8}$~gm~s$^{-1}$, where $R_\mathrm{10}$ is the radius of the disk in units of
$10^{10}$~cm (e.g. see Shafter et al. [1986], and references therein).
Values of $\dot M_\mathrm{crit}$ for our best-fitting models
are given in Tables~10 and~11. Most models
have mass accretion rates that are slightly above the critical value,
which is consistent with the lack of reported dwarf nova eruptions in
TT~Tri. However, the proximity of the inferred mass transfer rates to the
critical value suggests that it would not be surprising
if future observations revealed TT~Tri to exhibit standstills characteristic
of the Z~Cam stars.

In summary, the
disk luminosities and mass accretion rates for both the SD and TD models
are consistent
with what is expected for a $P=3.35$~hr novalike variable. Thus,
we are unable to distinguish between
the SD and TD classes of models for TT~Tri based on these criteria.

\subsection{The Distance to TT~Tri}

The distance to TT~Tri can be estimated
by comparing the observed brightness at mid-eclipse with
estimates of the absolute magnitude of the secondary star ($M_\mathrm{V,2}$),
and the fraction of light it contributes at mid-eclipse.
The fraction of light contributed by the secondary star
at mid-eclipse,
$f_2=f^\mathrm{s}_\mathrm{V}(\phi=0)/f^\mathrm{tot}_\mathrm{V}(\phi=0)$,
depends on the model considered, and
values of $f_2$ for each model
are given in Tables~10 and~11.
Generally, the secondary
star contributes between 3\% and 5\% of the light at mid-eclipse.
Given that $m_\mathrm{V}=16.42\pm0.25$ (see Table~2) at mid-eclipse,
we estimate the distance moduli $(m-M)_\mathrm{V}$ given in
Tables~10 and~11.
Within the errors of our measurements,
the observed and model $B-V$ colors agree well. Thus, $E(B-V)\simeq0$, and
there is no evidence
for any significant reddening along the line of sight to TT~Tri.

As a consistency check,
the distance to TT~Tri can also be determined from a comparison
of the out-of-eclipse magnitude with an estimate of the system's
absolute magnitude corrected for inclination. Warner (1987, 1995) gives
a correction factor for the conversion of absolute magnitude to
{\it apparent\/} absolute magnitude of
$\Delta M_\mathrm{V}(i)=-2.5~\mathrm{log} \left[(1+1.5~\mathrm{cos}~i)~\mathrm{cos}~i \right]$.
For our representative $q=0.6$ model, $i=72.6^{\circ}$, which leads to
a correction of $\sim0.9$~mag (where we have ignored
the small contribution of the disk rim).
If we assume that TT~Tri has an absolute
magnitude
$M_\mathrm{V}\simeq6.4\pm0.5$ (the mean of our two estimates from the previous
section),
and adopt $m_\mathrm{V}=15.61\pm0.14$ for TT~Tri out of eclipse (see Table~3),
we find $(m-M)_\mathrm{V}=8.3\pm0.5$,
and a corresponding distance of $\sim450$~pc.
Despite the uncertainties,
this value is consistent with our distance estimates
based on the observed light from the secondary star at mid-eclipse,
giving us confidence in our derived distance to TT~Tri.
The principal source of uncertainty in our distance estimates
arises from the adopted absolute magnitude of the secondary star,
which assumes that its properties follow the mean relations
for secondary stars in CVs determined by Smith \& Dhillon (1998).

\section{Conclusions}

We have performed the first multicolor eclipse study of the novalike
variable TT~Tri. The best-fitting model parameters are typical
of what is expected for a P=3.35~hr novalike variable, and are
relatively insensitive to the adopted mass ratio and rim thickness.
Our major conclusions are summarized below:

1) The mass ratio of TT~Tri is poorly constrained. As a result, we have
considered a wide range of possible values in our analysis:
$q=0.3$, $q=0.6$, and $q=0.9$.
We found that not only do the higher mass ratio models provide a better fit to
the data, they
are characterized by disks that are more typical of a system with
steady-state accretion (e.g. radii close to their tidal limit
with steep radial temperature profiles).
Nevertheless, acceptable model fits can be found for $q=0.3$,
and thus lower mass ratios cannot be ruled out by the data.
Fortunately, the values for the other fitting parameters are
very insensitive to the adopted mass ratio.

2) The radius of the accretion disk in TT~Tri extends to $\sim$50--60\%
of the distance to the inner Lagrangian point, which is
somewhat smaller than the maximum size allowed by tidal effects.
The disks in the higher mass ratio models fill a larger fraction of their
tidal radius, as expected if the disk is in a state of steady accretion.
In the $q=0.3$ models, the disk extends to only $\sim$70\% of the tidal
radius, while for the $q=0.6$ and $q=0.9$ models the disk
extends to $\sim$80\% and $\sim$90\% of the tidal
radius, respectively.
This result provides indirect evidence that the mass ratio
in TT~Tri may be closer to the maximum value allowed for stable accretion,
and that the white dwarf may have a relatively low mass
($M_1\simeq0.3-0.5$M$_{\odot}$).

3) As found in many other novalike systems, the best fitting
models of TT~Tri are not necessarily characterized by disks
with steep radial temperature profiles ($\alpha\simeq0.75$).
For example, SD models with values of $\alpha$ as low as 0.4
were found to provide acceptable
fits to the data, particularly in the cases where $q=0.3$.
It is worth recalling that
steady-state accretion does not necessarily require $\alpha=0.75$.
To begin with
our models fit brightness temperature, not the
the effective temperature used in eqn (2). Furthermore,
the effects of irradiation of the outer disk from hotter regions
near the white dwarf, and the potential cooling effects
of winds from the inner accretion disk, both of which would have the effect
of reducing the radial disk temperature gradient, have not been
included in our models. Despite these caveats,
it remains true that the smaller the value of $\alpha$,
the less confident we can be that the disk is in a state of steady accretion.

4) Unlike the case of V~Per,
where a thick disk rim was required to achieve temperature
profiles approaching $\alpha=0.75$, the thickness of the
rim had little effect in TT~Tri. However, significantly steeper
values of $\alpha$ were found when the inner disk was truncated, as one
would expect if the white dwarf is magnetic. A likely explanation
for the different behavior between V~Per and TT~Tri is that the
orbital inclination is significantly lower in TT~Tri. A lower inclination
will make the disk rim less important (e.g. Smak 1994), while making the
contribution of the inner disk relatively more important.
Despite the fact that the TD models are characterized
by significantly larger values of $\alpha$,
as just discussed, steady-state solutions were found
for the SD models as well. Thus, in our modeling,
we find no compelling evidence for
a truncated inner disk in TT~Tri.

5) We estimate that TT~Tri is characterized by a mass accretion rate,
$\dot M\sim 2\times10^{17}$~g~s$^{-1}$, which is just above
the critical rate for stable accretion. Given the proximity of the
accretion rate to the critical value, it is possible that future
observations of TT~Tri may reveal dwarf nova eruptions similar to
those seen in the Z~Cam stars, where the system fluctuates between
state of stable and unstable accretion (e.g. Shafter et al. 2005).
The TD models are characterized
by somewhat higher accretion rates, making them
more typical of novalike systems with stable accretion.

6) Finally,
by comparing the observed brightness of TT~Tri, both in and out of eclipse,
with properties of the secondary star and accretion disk,
we find that the system likely lies at a distance of $\sim400-500$~pc
from Earth.
The principle source of error in this estimate comes from uncertainty
in the assumed properties of the secondary star.




\acknowledgments
We thank Jerry Orosz for computing pictorial representations of the
system geometry near mid-eclipse, and an anonymous referee for numerous
suggestions that helped to improve our presentation.
Our eclipse model is based in part on
occultation kernel subroutines kindly provided by K. Horne.
This research was partially supported by a grant from NASA administered
by the American Astronomical Scoiety.

\clearpage

\begin{deluxetable}{ccccc}
\tablenum{1}
\tablewidth{0pt} 
\tablecolumns{5}
\tablecaption{Summary of Observations}
\tablehead{\colhead{} 					&	 
           \colhead{UT Time} 				& 
	   \colhead{Time Resolution\tablenotemark{a}} 	&
	   \colhead{Number of} 				&
	   \colhead{} 					\\ 
	   \colhead{UT Date} 				& 
	   \colhead{(start of observations)} 		&
	   \colhead{(sec)} 				& 
	   \colhead{Exposures} 				& 
	   \colhead{Filter}				}
\startdata
2005 Sep 05 &08:19:00.70 &22.06 &487 & $V$\\
2005 Oct 26 &05:42:32.10 &32.07 &222 & $B$\\
2005 Oct 27 &04:26:01.70 &32.07 &320 & $I$\\
2005 Nov 04 &03:48:41.00 &32.05 &354 & $B$\\
2005 Nov 04 &07:07:03.40 &32.05 &190 & $B$\\
2005 Nov 05 &02:33:00.30 &32.06 &750 & $R$\\
2005 Nov 06 &04:00:00.00 &32.06 &563 & $I$\\
2005 Nov 22 &01:29:00.70 &32.06 &177 & $R$\\
2005 Nov 22 &05:13:01.00 &32.05 &90  & $V$\\
2005 Nov 28 &01:38:01.00 &32.06 &151 & $B$\\
2005 Nov 28 &05:00:00.00 &32.15 &121 & $B$\\
2005 Nov 29 &03:55:30.80 &32.06 &278 & $I$\\
2005 Nov 29 &06:34:00.40 &32.05 &300 & $R$\\
2005 Nov 30 &03:40:30.50 &32.07 &500 & $V$\\
2005 Nov 30 &08:08:23.30 &32.05 &66  & $V$\\
2005 Dec 01 &01:58:31.00 &32.06 &350 & $V$\\
2005 Dec 01 &05:17:00.40 &32.06 &360 & $V$\\
2005 Dec 21 &02:27:00.00 &32.06 &170 & $B$\\
2005 Dec 22 &02:12:00.90 &32.06 &171 & $V$\\
2005 Dec 22 &05:06:00.80 &32.06 &204 & $R$\\
2005 Dec 24 &02:05:00.50 &32.06 &505 & $R$\\
\enddata
\label{obser}
\tablenotetext{a}{Mean time interval between exposures (integration time plus 
readout time)}
\end{deluxetable}

\clearpage

\begin{deluxetable}{lccc}
\tablenum{2}
\tablecaption{Mean Magnitudes and Colors}
\tablewidth{0pt}
\tablehead{
& \colhead{Effective} & &  \\
\colhead{Photometric} & \colhead{Wavelength} & & \\
\colhead{Parameter} & \colhead{(\AA)} & \colhead{Out of Eclipse\tablenotemark{a}} & \colhead{Primary Minimum}
} 
\startdata
$B$\dots\dots\dots & 4386 & $15.69\pm0.15$ & $16.59\pm0.30$ \\
$V$\dots\dots\dots & 5508 & $15.61\pm0.14$ & $16.42\pm0.25$ \\
$R$\tablenotemark{b}\dots\dots\dots & 6518 & ----- & ----- \\
$I$\tablenotemark{b}\dots\dots\dots & 8239 & ----- & ----- \\
\\
$B-V$\dots & \dots & $0.08\pm0.20$ & $0.17\pm0.39$ \\

\enddata
\tablenotetext{a}{The mean out-of-eclipse magnitude
was computed with the pre-eclipse level weighted twice
that of the post-eclipse level, as described in the text.}
\tablenotetext{b}{The $R$ and $I$ photometry was not calibrated.}
\end{deluxetable}

\clearpage

\begin{deluxetable}{lccr}
\tablenum{3}
\tablecaption{Eclipse Timings}
\tablewidth{0pt}
\tablehead{	\colhead{HJD (mid-eclipse)} 	& 
		\colhead{Cycle Number}		& 
		\colhead{}			& 
		\colhead{$O-C$} 		\\
		\colhead{(2,400,000+)} 		& 
		\colhead{$(E)$} 		& 
		\colhead{Filter} 		& 
		\colhead{($\times10^3$~day)}	} 
\startdata
53618.9531\dots &  0.   & $V$ &  $ 0.501730$  \\
53669.7801\dots &  364. & $B$ &  $-0.315169$  \\
53670.7576\dots &  371. & $I$ &  $-0.222475$  \\
53678.7171\dots &  428. & $B$ &  $-0.083822$  \\
53678.8564\dots &  429. & $B$ &  $-0.344723$  \\
53679.6950\dots &  435. & $R$ &  $ 0.405872$  \\
53679.8339\dots &  436. & $R$ &  $-0.361029$  \\
53680.8110\dots &  443. & $I$ &  $-0.755334$  \\
53696.5913\dots &  556. & $R$ &  $ 0.632873$  \\
53696.7305\dots &  557. & $V$ &  $ 0.213972$  \\
53702.5952\dots &  599. & $B$ &  $ 0.120137$  \\
53702.7347\dots &  600. & $B$ &  $-0.058764$  \\
53703.7122\dots &  607. & $I$ &  $ 0.028931$  \\
53703.8519\dots &  608. & $R$ &  $ 0.126030$  \\
53704.6893\dots &  614. & $V$ &  $-0.314375$  \\
53704.8296\dots &  615. & $V$ &  $ 0.352724$  \\
53705.6671\dots &  621. & $V$ &  $ 0.012319$  \\
53705.8067\dots &  622. & $V$ &  $-0.043582$  \\
53725.6346\dots &  764. & $B$ &  $-0.555499$  \\
53726.6128\dots &  771. & $V$ &  $ 0.206196$  \\
53726.7523\dots &  772. & $R$ &  $ 0.093295$  \\
53728.7075\dots &  786. & $R$ &  $ 0.360683$  \\
\enddata
\label{time}
\end{deluxetable}

\clearpage

\begin{deluxetable}{lcc}
\tablecaption{Model Input Parameters}
\tablenum{4}
\tablewidth{0pt}
\tablehead{
\colhead{Parameter} & \colhead{Definition} & \colhead{Value\tablenotemark{a}}
} 
\startdata
$P$\dots\dots & Orbital Period, $P$ & 3.35~hr \\
$q$\dots\dots & Mass ratio, $M_2/M_1$ & 0.3, 0.6, 0.9 \\
$\Delta\phi$\dots\dots & Eclipse phase width & 0.035\\
$i$\dots\dots & Orbital inclination & computed \\
$\alpha$\dots\dots & Disk temperature parameter & variable (0.25--0.95)\\
$R_{\mathrm{d}}$\dots\dots & Disk radius & variable (0.35--0.70)\\
$R_{\mathrm{in}}$\dots\dots & Inner disk radius & $R_1$, $0.05R_{\mathrm{L1}}$, $0.10R_{\mathrm{L1}}$\\
$h_{\mathrm{r}}$\dots\dots & Disk rim parameter& 0.04, 0.08 \\
$T_{\mathrm{d}}$\dots\dots & Temperature of disk perimeter & variable (5000--12000K)\\
$R_1$\dots\dots & Radius of white dwarf & computed\\
$R_2$\dots\dots & Radius of secondary star & computed\\
$T_1$\dots\dots & Temperature of white dwarf & variable (10000--80000K)\\
$T_2$\dots\dots & Temperature of secondary star & 3400~K\\
$\chi_\mathrm{s}$\dots\dots & Bright spot temperature factor & variable (1.0--2.4)\\
$R_s$\dots\dots & Bright spot radius & 0.2$R_{\mathrm{d}}$ \\
\enddata
\tablenotetext{a}{Where given explicitly,
the values are fixed and not varied during the
fitting procedure. The values of ``variable" parameters are determined
by the model during the fitting procedure, while the
``computed" values are calculated by the model for a given mass ratio.}
\end{deluxetable}

\clearpage

\begin{deluxetable}{lccc}
\tablecaption{Binary Parameters}
\tablenum{5}
\tablewidth{0pt}
\tablehead{
\colhead{Parameter} & \colhead{$q=0.3$} & \colhead{$q=0.6$} & \colhead{$q=0.9$}}
\startdata
$i(^\circ)$\dots\dots\dots    & 76.1 & 72.6 & 70.4\\
$a(R_{\odot})$\dots\dots\dots & 1.17 & 1.01 & 0.95\\
$R_{\mathrm{L1}}(a)$\dots\dots\dots  & 0.62 & 0.55 & 0.51 \\
$R_1(10^{-2}~R_{\odot})$\dots & 0.94 & 1.48 & 1.75\\
$M_1(M_{\odot}$)\dots\dots    & 0.86 & 0.45 & 0.31\\
$R_2(R_{\odot}$)\dots\dots    & 0.33 & 0.34 & 0.35\\
$M_2(M_{\odot}$)\dots\dots    & 0.26 & 0.27 & 0.28\\
\enddata
\label{param}
\end{deluxetable}

\clearpage

\begin{deluxetable}{lcccccccc}
\rotate
\tablecaption{Standard Disk Model Grid Solutions}
\tablenum{6}
\tablewidth{0pt}
\tablehead{&\colhead{$q=0.3$} &&\colhead{} & \colhead{$q=0.6$} &&\colhead{}& \colhead{$q=0.9$} & \\
\colhead{Parameter} & \colhead{$h_{\mathrm{r}}=0.04$} & 
\colhead{$h_{\mathrm{r}}=0.08$} & \colhead{}& \colhead{$h_{\mathrm{r}}=0.04$} &
\colhead{$h_{\mathrm{r}}=0.08$} & \colhead{}& \colhead{$h_{\mathrm{r}}=0.04$}&
\colhead{$h_{\mathrm{r}}=0.08$}} 
\startdata
$R_{\mathrm{d}}(a)$           & 0.31 & 0.31 & & 0.30 & 0.30 & & 0.28 & 0.28 \\
$R_{\mathrm{d}}(a)_{\mathrm{lim}}$\tablenotemark{a} & 0.46 & 0.46 & & 0.38 &
0.38 & & 0.32 & 0.32 \\ 
$R_{\mathrm{d}}(R_{\mathrm{L1}})$      & 0.50 & 0.50 &&0.55 & 0.55 & &
0.55&0.55  \\
$\alpha$           & 0.45  & 0.55& &0.55 & 0.55 & & 0.65 &0.65\\
$T_1$($10^3$~K)    & 10.0 &  20.0 &&10.0 & 20.0 & & 10.0  &10.0\\
$T_{\mathrm{d}}$($10^3$~K)    & 7.0  &  8.0 &&7.0  & 7.0 & & 7.0 &7.0\\
$\chi_\mathrm{s}$              &1.2  & 1.2 &&1.6 &1.4&  & 1.6 &1.4\\
$(B-V)_{\circ}$    & 0.21  &  0.10& &0.14 & 0.15&  & 0.09 &0.11\\
$\chi_{\nu,{\mathrm{min}}}^2$     & 1.67  &  1.86&&  1.15 & 1.36&& 1.15 &1.30 \\
\enddata
\tablenotetext{a}{The limiting tidal radius of the disk computed using
eqn~(2.61) of Warner (1995).}
\end{deluxetable}

\clearpage

\begin{deluxetable}{lcccccccc}
\rotate
\tablecaption{Truncated Disk Model Grid Solutions}
\tablenum{7}
\tablewidth{0pt}
\tablehead{&\colhead{$q=0.3$} &&\colhead{} & \colhead{$q=0.6$} &&\colhead{}& \colhead{$q=0.9$} & \\
\colhead{Parameter} & \colhead{$R_\mathrm{in}=0.05R_\mathrm{L1}$} & 
\colhead{$R_\mathrm{in}=0.10R_\mathrm{L1}$} & \colhead{}& \colhead{$R_\mathrm{in}=0.05R_\mathrm{L1}$} &
\colhead{$R_\mathrm{in}=0.10R_\mathrm{L1}$} & \colhead{}& \colhead{$R_\mathrm{in}=0.05R_\mathrm{L1}$}&
\colhead{$R_\mathrm{in}=0.10R_\mathrm{L1}$}} 
\startdata
$R_{\mathrm{d}}(a)$           & 0.31 & 0.31 & & 0.30 & 0.30 & & 0.28 & 0.28 \\
$R_{\mathrm{d}}(a)_{\mathrm{lim}}$\tablenotemark{a} & 0.46 & 0.46 & & 0.38 &
0.38 & & 0.32 & 0.32 \\ 
$R_{\mathrm{d}}(R_{\mathrm{L1}})$      & 0.55 & 0.50 &&0.55 & 0.55 & &
0.55&0.55  \\
$\alpha$           & 0.75  & 0.85& &0.65 & 0.85 & & 0.75 &0.85\\
$T_1$($10^3$~K)    & 40.0 &  80.0 &&20.0 & 30.0 & & 10.0  &20.0\\
$T_{\mathrm{d}}$($10^3$~K)    & 8.0  &  8.0 &&7.0  & 7.0 & & 7.0 &7.0\\
$\chi_\mathrm{s}$              &1.4  & 1.2 &&1.6 &1.6&  & 1.6 &1.6\\
$(B-V)_{\circ}$    & 0.02  &  0.04& &0.11 & 0.07&  & 0.07 &0.09\\
$\chi_{\nu,{\mathrm{min}}}^2$     & 1.49 & 1.78&&  1.14 & 1.48&& 1.15 &1.53 \\
\enddata
\tablenotetext{a}{The limiting tidal radius of the disk computed using
eqn~(2.61) of Warner (1995).}
\end{deluxetable}

\clearpage

\begin{deluxetable}{lcccccccc}
\rotate
\tablecaption{SD Model: Distribution Mean Solutions}
\tablenum{8}
\tablewidth{0pt}
\tablehead{&\colhead{$q=0.3$} &&\colhead{} & \colhead{$q=0.6$} &&\colhead{}& \colhead{$q=0.9$} & \\
\colhead{Parameter} & \colhead{$h_{\mathrm{r}}=0.04$} & 
\colhead{$h_{\mathrm{r}}=0.08$} & \colhead{}& \colhead{$h_{\mathrm{r}}=0.04$} &
\colhead{$h_{\mathrm{r}}=0.08$} & \colhead{}& \colhead{$h_{\mathrm{r}}=0.04$}&
\colhead{$h_{\mathrm{r}}=0.08$}} 
\startdata
$<\chi_\nu^2>$  &  1.79 & 2.01 && 1.41 & 1.59 && 1.47 &1.55\\
$R_{\mathrm{d}}(R_{\mathrm{L1}})$  & $0.52\pm0.03$ &  $0.48\pm0.03$ && $0.57\pm0.04$ & $0.54\pm0.03$ && $0.55\pm0.05$ & $0.55\pm0.04$ \\
$\alpha$           & $0.59\pm0.10$ &  $0.55\pm0.12$  && $0.66\pm0.11$  &  $0.68\pm0.12$ && $0.66\pm0.12$ & $0.71\pm0.13$ \\
$T_1$($10^3$~K)    & $32.9\pm13.1$ & $37.7\pm17.4$ && $16.7\pm6.26$ & $23.8\pm10.5$ && $12.9\pm3.82$ & $17.2\pm5.73$ \\
$T_{\mathrm{d}}$($10^3$~K)    & $9.08\pm1.17$ & $9.23\pm1.49$ && $8.06\pm1.14$ & $9.05\pm1.42$ && $7.58\pm0.97$ & $8.47\pm1.24$ \\
$\chi_\mathrm{s}$              & $1.39\pm0.14$ & $1.18\pm0.06$ && $1.62\pm0.19$ & $1.45\pm0.14$ && $1.58\pm0.20$ & $1.47\pm0.16$ \\
$(B-V)_{\circ}$    & $-0.03\pm0.10$ & $0.00\pm0.12$ && $-0.01\pm0.11$ & $-0.05\pm0.11$ && $0.01\pm0.11$ & $-0.03\pm0.11$  \\
\enddata
\end{deluxetable}

\clearpage

\begin{deluxetable}{lcccccccc}
\rotate
\tablecaption{TD Model: Distribution Mean Solutions}
\tablenum{9}
\tablewidth{0pt}
\tablehead{&\colhead{$q=0.3$} &&\colhead{} & \colhead{$q=0.6$} &&\colhead{}& \colhead{$q=0.9$} & \\
\colhead{Parameter} & \colhead{$R_{\mathrm{in}}=0.05R_\mathrm{L1}$} & 
\colhead{$R_{\mathrm{in}}=0.10R_\mathrm{L1}$} & \colhead{}& \colhead{$R_{\mathrm{in}}=0.05R_\mathrm{L1}$} &
\colhead{$R_{\mathrm{in}}=0.10R_\mathrm{L1}$} & \colhead{}& \colhead{$R_{\mathrm{in}}=0.05R_\mathrm{L1}$}&
\colhead{$R_{\mathrm{in}}=0.10R_\mathrm{L1}$}}
\startdata
$<\chi_\nu^2>$  &  1.60 & 1.98 && 1.30 & 1.69 && 1.40 &1.79\\
$R_{\mathrm{d}}(R_{\mathrm{L1}})$  & $0.54\pm0.05$ &  $0.49\pm0.03$ && $0.58\pm0.04$ & $0.56\pm0.05$ && $0.57\pm0.04$ & $0.57\pm0.05$ \\
$\alpha$           & $0.76\pm0.11$ &  $0.81\pm0.11$  && $0.78\pm0.12$  &  $0.87\pm0.09$ && $0.77\pm0.12$ & $0.87\pm0.10$ \\
$T_1$($10^3$~K)    & $42.8\pm12.4$ & $61.4\pm10.8$ && $19.5\pm6.88$ & $40.1\pm10.8$ && $13.1\pm4.62$ & $24.1\pm6.02$ \\
$T_{\mathrm{d}}$($10^3$~K)    & $8.49\pm0.97$ & $7.72\pm0.84$ && $7.66\pm0.99$ & $7.51\pm1.12$ && $7.28\pm0.94$ & $6.97\pm1.07$ \\
$\chi_\mathrm{s}$              & $1.44\pm0.18$ & $1.24\pm0.11$ && $1.67\pm0.18$ & $1.57\pm0.18$ && $1.65\pm0.20$ & $1.57\pm0.22$ \\
$(B-V)_{\circ}$    & $-0.05\pm0.08$ & $0.03\pm0.07$ && $-0.03\pm0.09$ & $-0.01\pm0.08$ && $-0.01\pm0.09$ & $0.02\pm0.07$  \\
\enddata
\end{deluxetable}

\clearpage

\begin{deluxetable}{lcccccccc}
\rotate
\tablenum{10}
\tablecaption{SD Models: Derived Properties}
\tablewidth{0pt}
\tablehead{&\colhead{$q=0.3$} &&\colhead{} & \colhead{$q=0.6$} &&\colhead{}& \colhead{$q=0.9$} & \\
\colhead{Parameter} & \colhead{$h_{\mathrm{r}}=0.04$} & 
\colhead{$h_{\mathrm{r}}=0.08$} & \colhead{}& \colhead{$h_{\mathrm{r}}=0.04$} &
\colhead{$h_{\mathrm{r}}=0.08$} & \colhead{}& \colhead{$h_{\mathrm{r}}=0.04$}&
\colhead{$h_{\mathrm{r}}=0.08$}}
\startdata
$\eta_\mathrm{d,s}$ & 159& 274&& 126 & 125 && 103 & 102 \\
$M_\mathrm{V}$ & 6.7 & 6.2 && 7.0 & 7.0 && 7.2 & 7.2 \\
$\dot M_{17}~(\mathrm{g~s}^{-1})$ & 1.9 & 3.3 && 2.4 & 2.4 && 2.4 & 2.4 \\
$\dot M_\mathrm{crit,17}~(\mathrm{g~s}^{-1})$ & 1.3 & 1.3 && 1.5 & 1.5 && 1.4 & 1.4 \\
$f_2$  & 0.050 & 0.027 && 0.053 & 0.048 && 0.060 & 0.055 \\ 
$(m-M)_\mathrm{V}$ & 8.2 & 8.9 && 8.1 & 8.2 && 8.0 & 8.1 \\
$d~(\mathrm{pc})$ & 430 & 590 && 420 & 440 && 390 & 410 \\ 
\enddata
\end{deluxetable}

\clearpage

\begin{deluxetable}{lcccccccc}
\rotate
\tablenum{11}
\tablecaption{TD Models: Derived Properties}
\tablewidth{0pt}
\tablehead{&\colhead{$q=0.3$} &&\colhead{} & \colhead{$q=0.6$} &&\colhead{}& \colhead{$q=0.9$} & \\
\colhead{Parameter} & \colhead{$R_\mathrm{in}=0.05R_\mathrm{L1}$} & 
\colhead{$R_\mathrm{in}=0.10R_\mathrm{L1}$} & \colhead{}& \colhead{$R_\mathrm{in}=0.05R_\mathrm{L1}$} &
\colhead{$R_\mathrm{in}=0.10R_\mathrm{L1}$} & \colhead{}& \colhead{$R_\mathrm{in}=0.05R_\mathrm{L1}$}&
\colhead{$R_\mathrm{in}=0.10R_\mathrm{L1}$}}
\startdata
$\eta_\mathrm{d,s}$ & 388& 249&& 133 & 131 && 112 & 93 \\
$M_\mathrm{V}$ & 5.8 & 6.3 && 6.9 & 6.9 && 7.1 & 7.3 \\
$\dot M_{17}~(\mathrm{g~s}^{-1})$ & 5.3 & 5.0 && 2.7 & 3.3 && 2.5 & 3.1 \\
$\dot M_\mathrm{crit,17}~(\mathrm{g~s}^{-1})$ & 1.7 & 1.3 && 1.5 & 1.5 && 1.4 & 1.4 \\
$f_2$  & 0.021 & 0.029 && 0.050 & 0.045 && 0.055 & 0.058 \\ 
$(m-M)_\mathrm{V}$ & 9.1 & 8.8 && 8.2 & 8.3 && 8.1 & 8.0 \\
$d~(\mathrm{pc})$ & 660 & 570 && 430 & 460 && 410 & 400 \\ 
\enddata
\end{deluxetable}

\clearpage

\begin{figure}
\epsscale{0.80}
\plotone{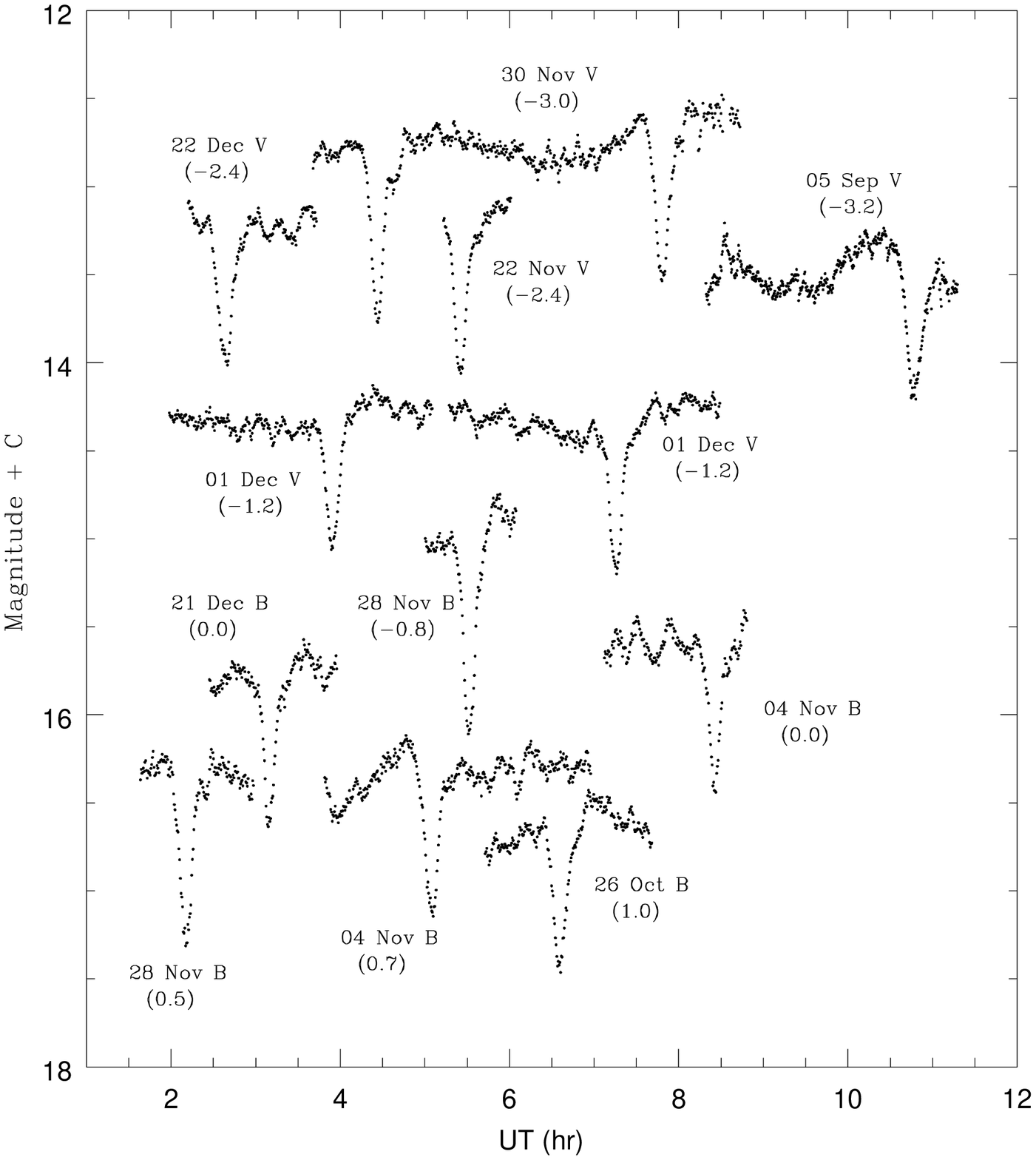}
\caption{The $B$ and $V$ light curves of TT~Tri. For clarity of
          presentation, the data have been
          offset by addition of the constants given in parentheses
          under each date.}
\end{figure}

\clearpage

\begin{figure}
\epsscale{0.80}
\plotone{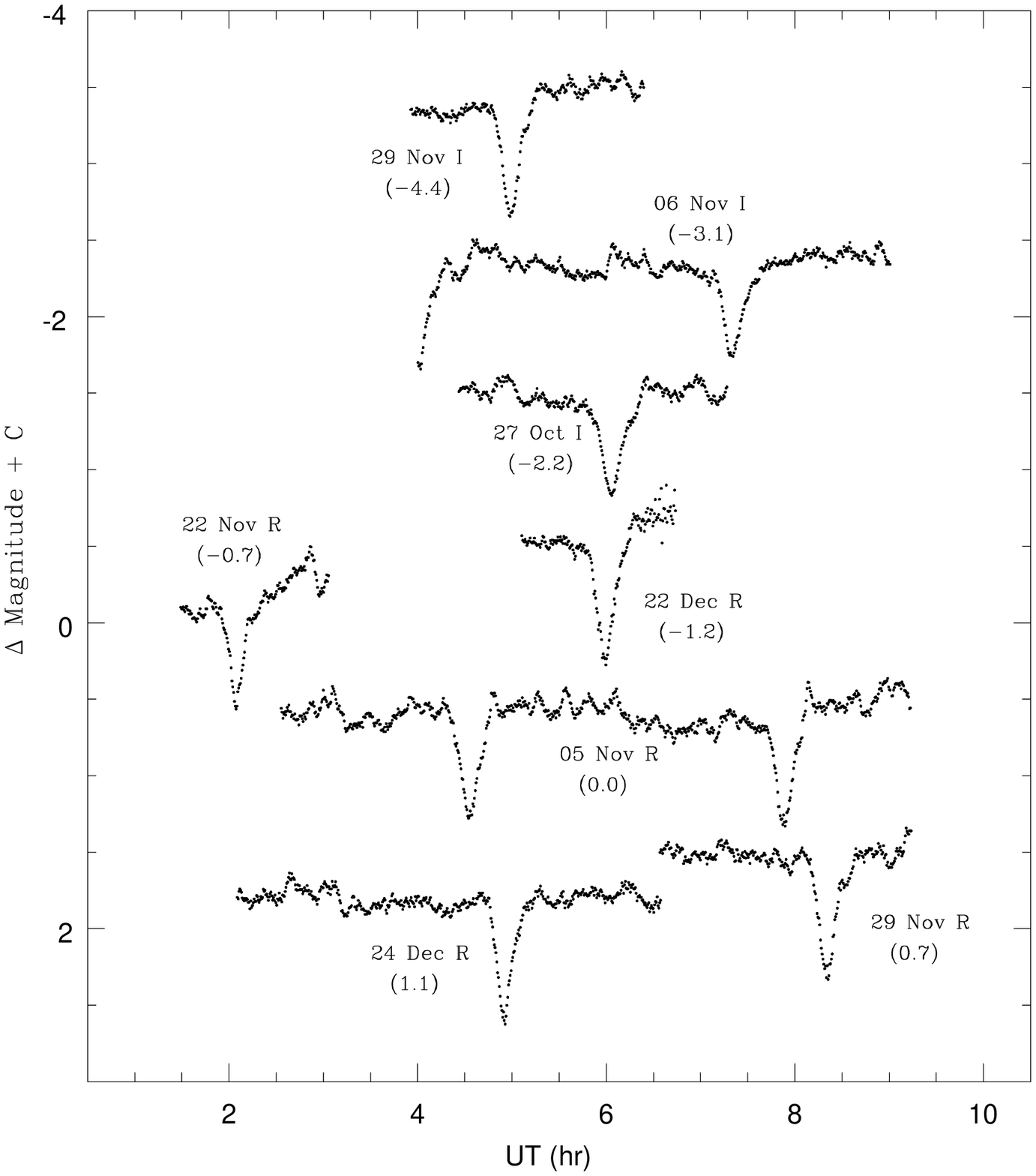}
\caption{The $R$ and $I$ light curves of TT~Tri. For clarity of
          presentation, the data have been
          offset as in Fig. 1 by addition of the constants given in parentheses
          under each date.}
\end{figure}

\begin{figure}
\epsscale{0.80}
\plotone{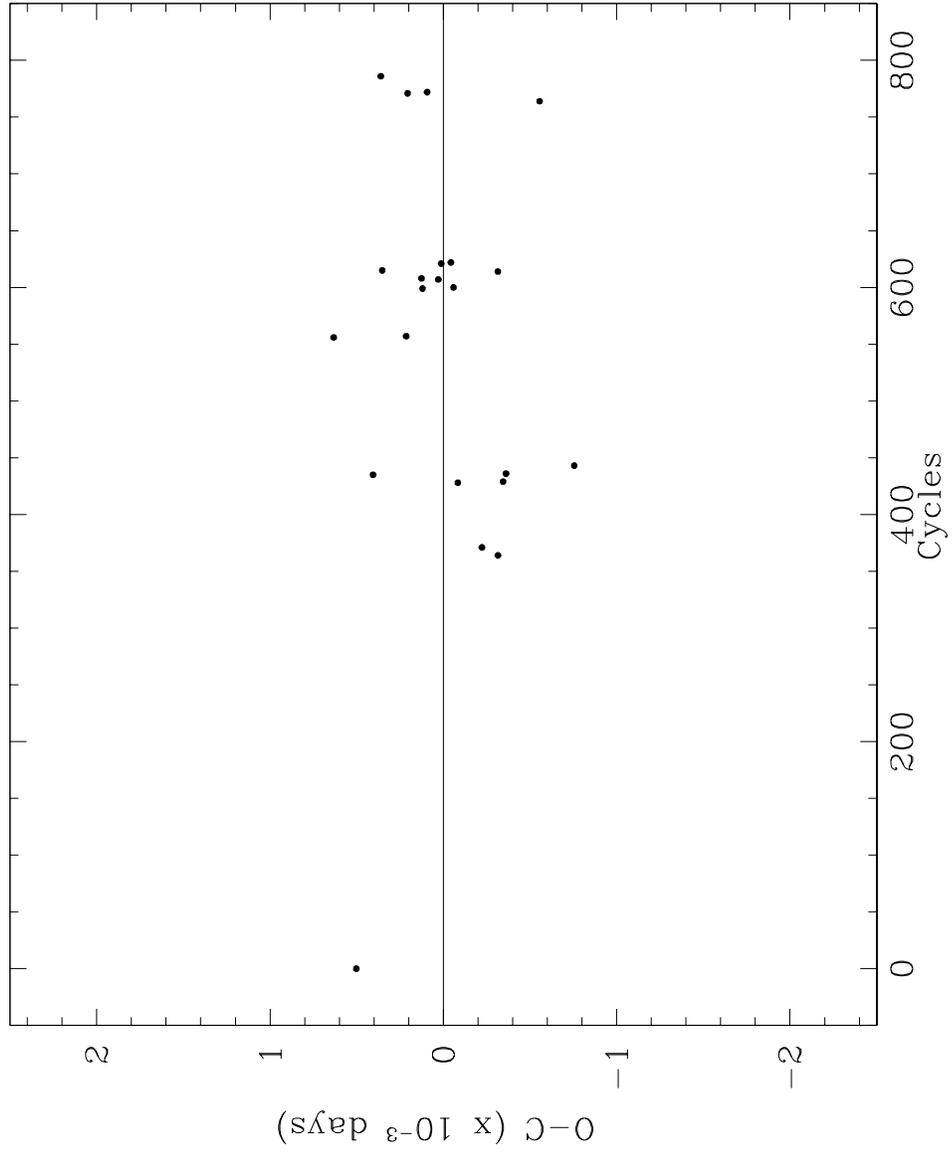}
\caption{The residuals of the the observed times of mid-eclipse
          with respect to eqn~(1) is plotted as a function of
          cycle number. All timings are within $\sim 1$~min of those
          predicted by the ephemeris.}
\end{figure}

\clearpage

\begin{figure}
\epsscale{0.80}
\plotone{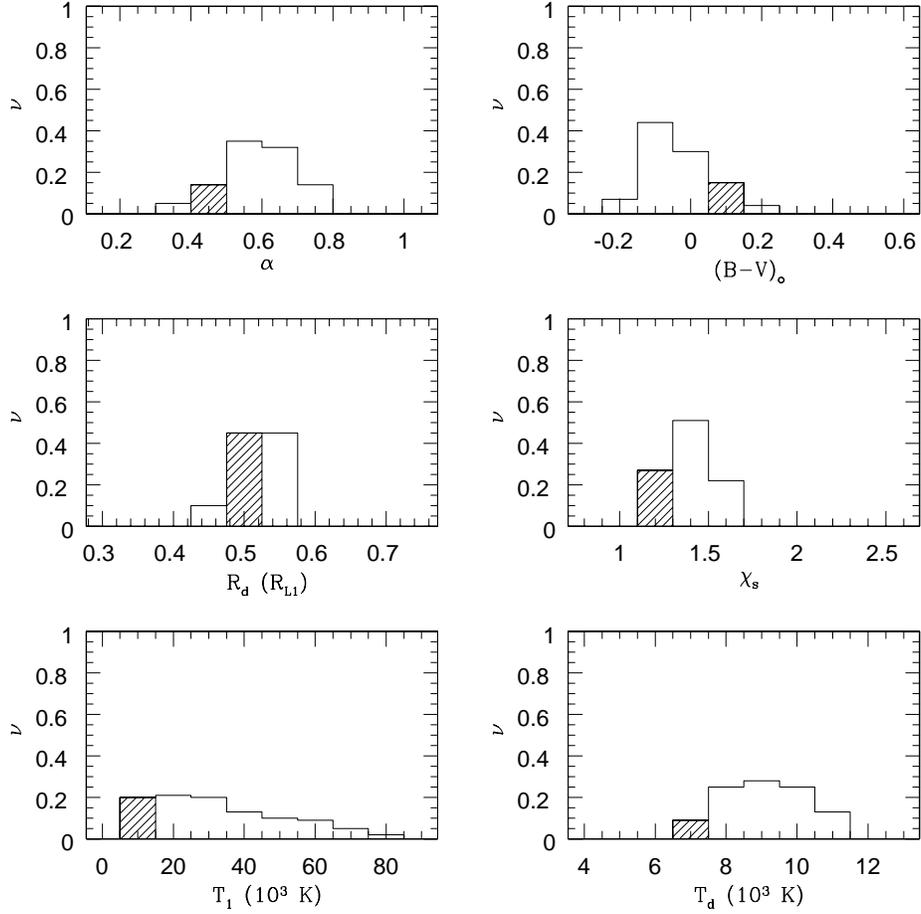}
\caption{The $q=0.3, h_{\mathrm{r}}=0.04$
          frequency distributions for each fitting parameter.
          The model $B-V$ value has also been included.
          The range of each parameter has been chosen
          to include all combinations of parameters that result in
          acceptable fits of the top 100 best-fitting models to the data.
          The cross-hatched regions indicate the parameter values
          that produce the optimum fit to the data (Table~6).}
\end{figure}

\clearpage

\begin{figure}
\epsscale{0.80}
\plotone{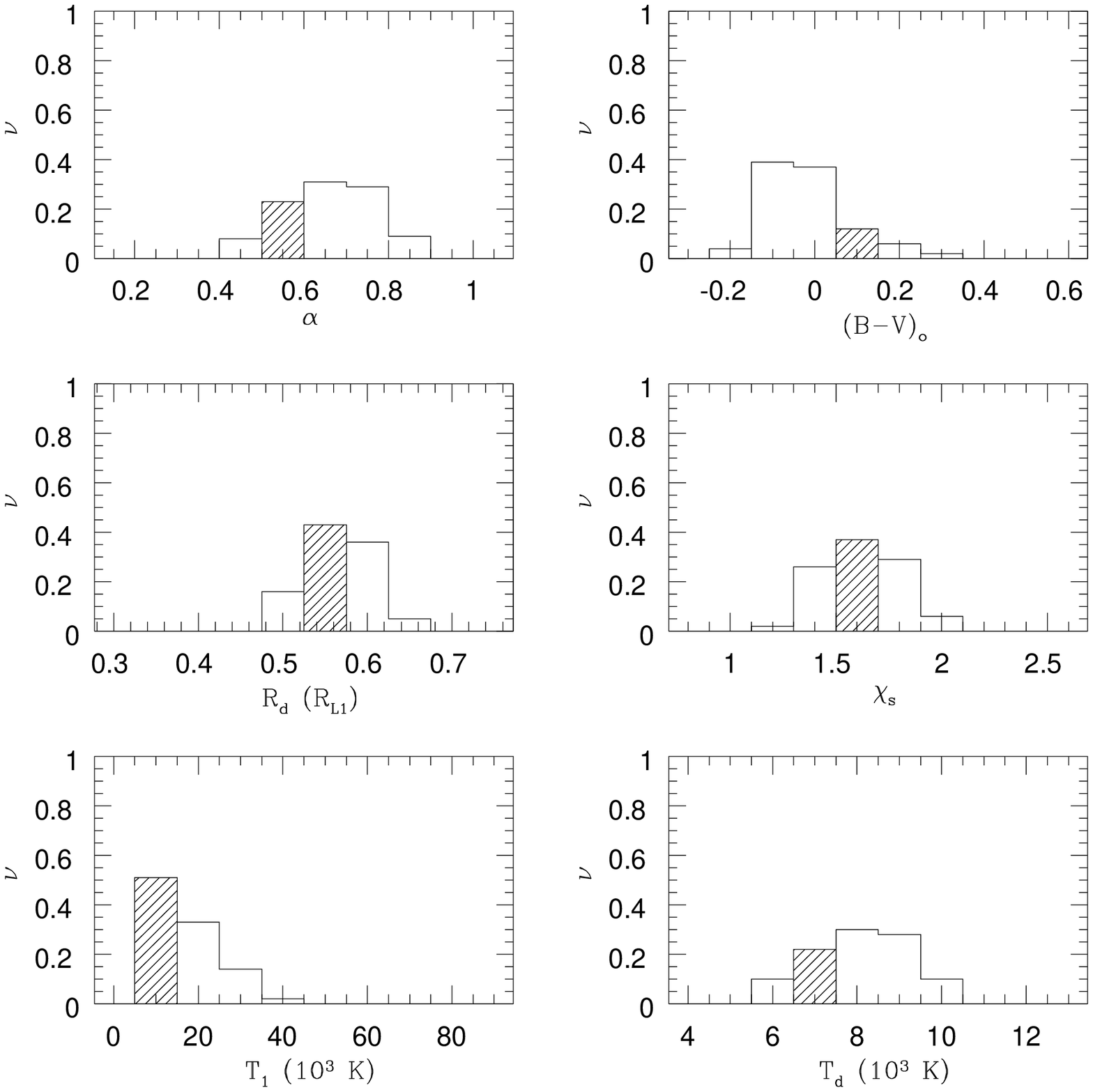}
\caption{The same as Figure~4, but for $q=0.6$}
\end{figure}

\clearpage

\begin{figure}
\epsscale{0.80}
\plotone{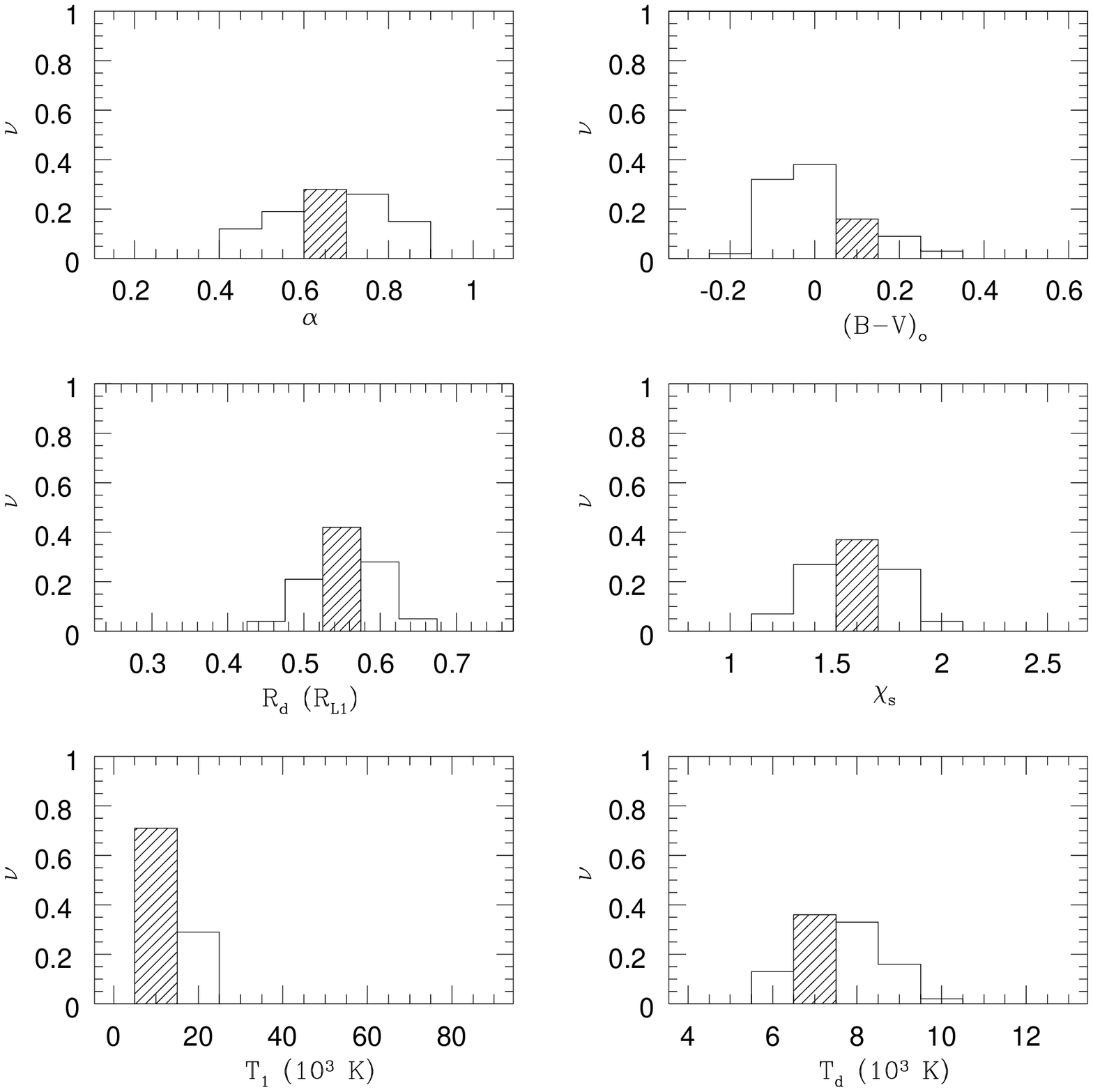}
\caption{The same as Figure~4, but for $q=0.9$}
\end{figure}

\clearpage

\begin{figure}
\epsscale{0.80}
\plotone{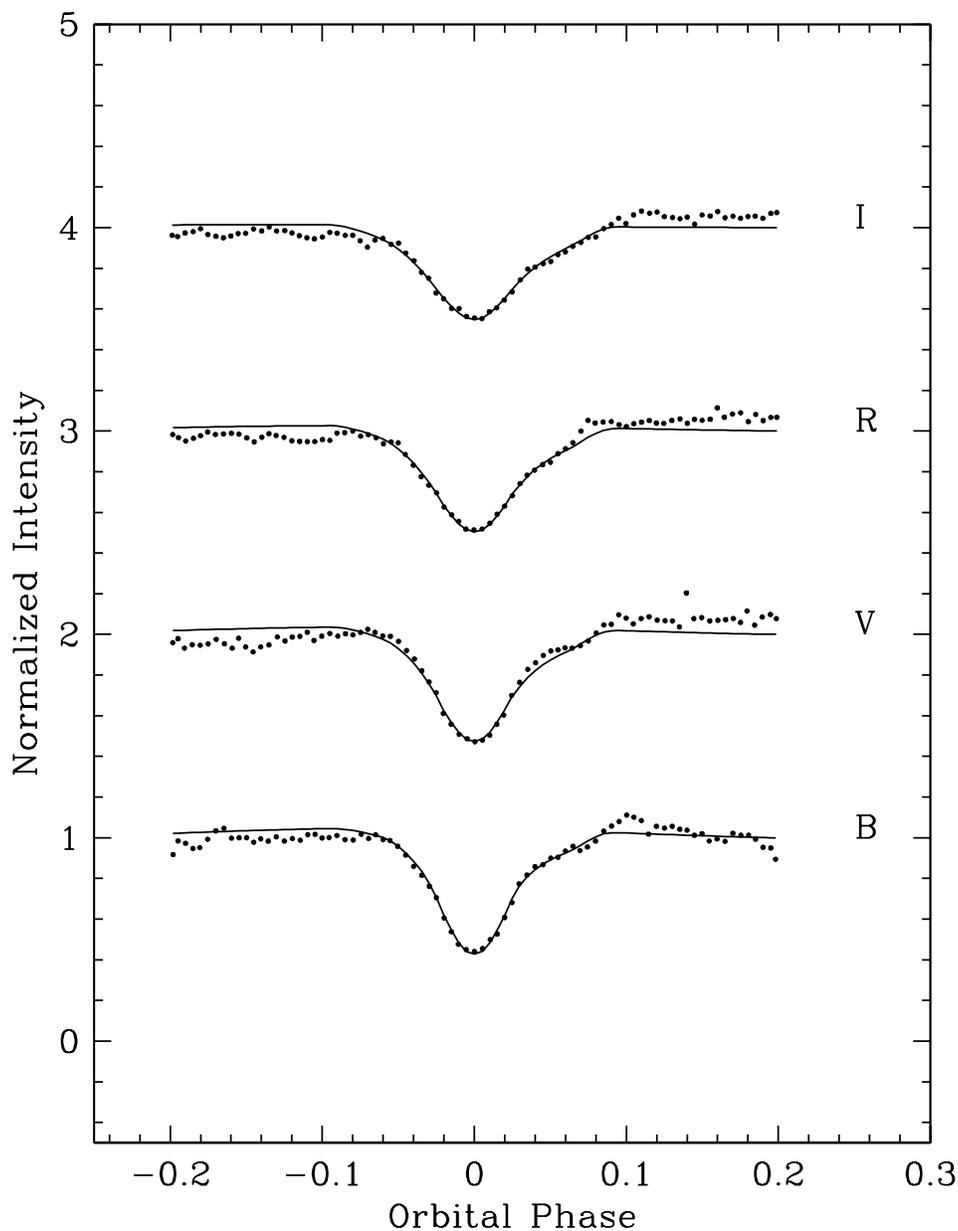}
\caption{The best-fitting eclipse profiles (solid lines) for
          the $q=0.6, h_\mathrm{r}=0.04$ SD
          model are plotted
          together with the observed data. The data have been converted to
          relative intensity and normalized to unity outside of eclipse,
          as described in the text.
          For clarity of presentation the $V$, $R$, and $I$-band
          light curves and models have been shifted
          upward by constant offsets of 1, 2, and 3, respectively.}
\end{figure}

\clearpage

\begin{figure}
\epsscale{0.80}
\plotone{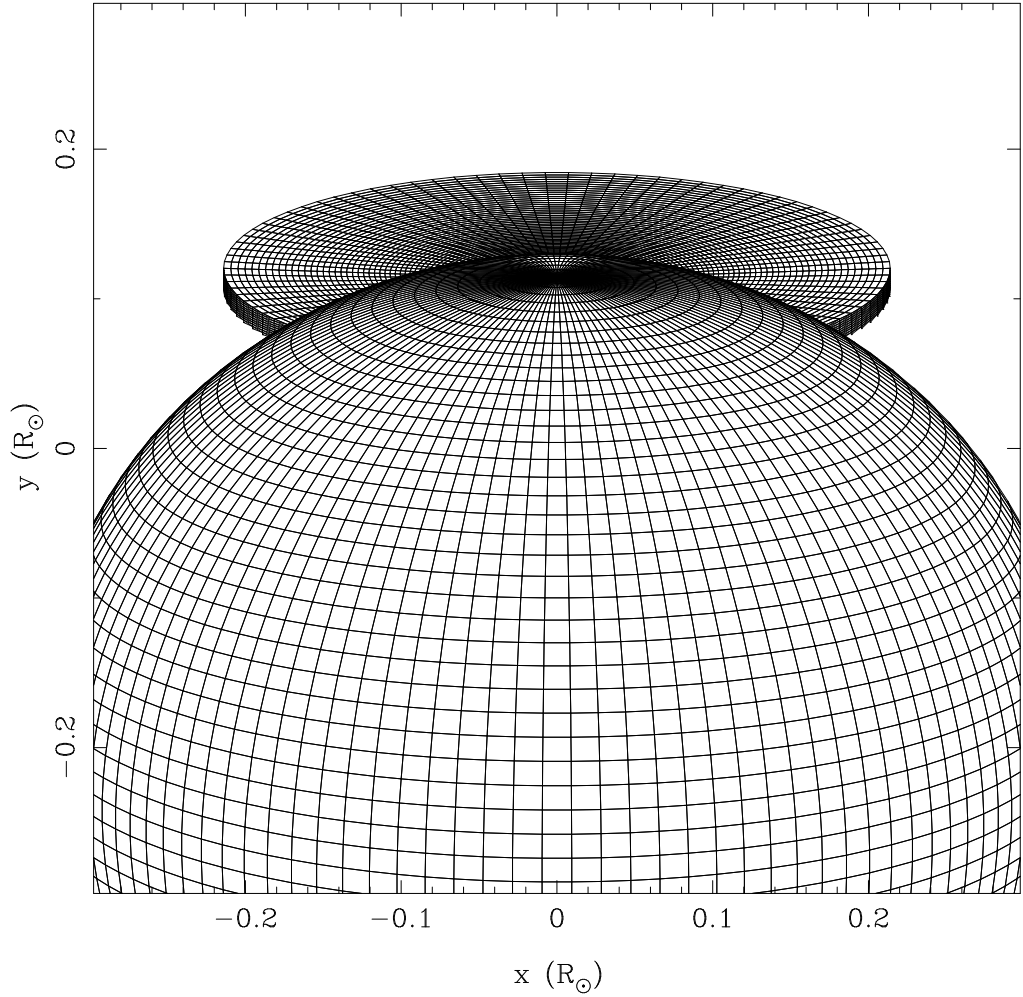}
\caption{A schematic representation of the TT~Tri geometry at mid-eclipse
          for the $q=0.6, i=72.6^{\circ}$ model. Note that the
          accretion disk is never completely eclipsed, and the white
          dwarf is barely eclipsed.}
\end{figure}

\clearpage

\begin{figure}
\epsscale{0.80}
\plotone{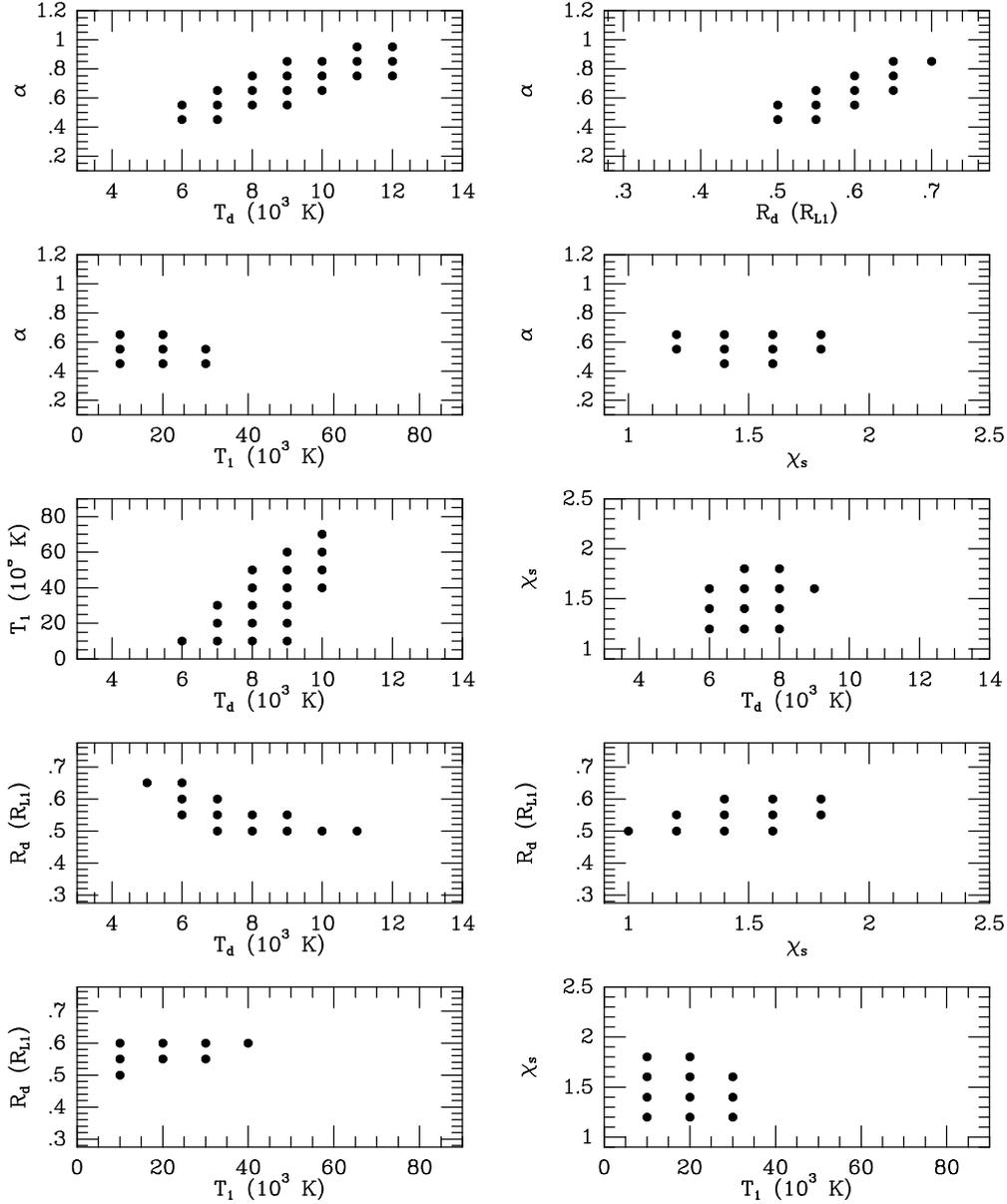}
\caption{The correlations between pairs of model parameters for the
          $q=0.6, h_\mathrm{r}=0.04$ SD model are shown
          for each of the 10 possible pairings of the five fitting
          parameters.
          In each case a range of model solutions ($\chi_\nu^2 < 2.0$)
          for each pair of parameters is plotted, while the remaining
          three parameters are held fixed at their optimum values (Table~6).}
\end{figure}

\end{document}